\documentclass[reprint,prd,amsmath,showpacs,aps]{revtex4-1}
\usepackage{hyperref}
\usepackage{bm}

\begin{document}
\title{Electroweak Corrections to the True Muonium Hyperfine Splitting}

\author{Henry Lamm}
\email{hlammiv@asu.edu}

\affiliation{Physics Department, Arizona State University, Tempe, AZ 85287}
\date{\today}

\begin{abstract}
In contrast to other atomic systems, in true muonium ($\mu^+\mu^-$) the leading-order $Z$ boson corrections to the hyperfine splitting are shown to be experimentally accessible in the near future.  This contribution ($-109$ MHz) constitutes a necessary contribution to a full $\mathcal{O}(m\alpha^7)$ calculation of the true muonium hyperfine splitting.  This calculation would enable a number of possible to the muon problem to be constrained.  Additionally, we compute the general expression for a pseudovector coupling to particle-antiparticle bound states at leading order, including the annihilation channel. 
\end{abstract}
\pacs{36.10.Ee, 31.30.jr, 32.10.Fn, 24.80.+y}

\maketitle
\section{Introduction}
\label{sec:1}
The current state of flavor physics might be described as having a ``muon problem''.  Several muon observables have been shown to have varying levels of disagreement with Standard Model calculations.  In Table~\ref{tab2} are listed a few of the most prominent signals that are in tension with expectations.
\begin{table}[ht]
\begin{center}
 \begin{tabular}{c c c c}
  \hline\hline
  Obs. &Discrep. & Ref.\\
  \hline
     $(g-2)_\mu$&$2.9\sigma$& \cite{PhysRevD.73.072003} \\
     $r_P$ from $\mu^-p^+$&$7\sigma$&\cite{Antognini:2012ofa}\\
     $B^+\rightarrow K^+\mu^+\mu^-$	& $2.6\sigma$ &\cite{Aaij:2014ora}\\
    $h\rightarrow \mu\tau$	& $2.5\sigma$&\cite{CMS:2014hha}\\
  \hline\hline
 \end{tabular}
\end{center}
\caption{\label{tab2} Set of prominent muon signals that have disagreements with theoretical expectations.}
 \end{table}
 
A strong candidate for shedding light on the muon problem is the bound state $(\mu^+\mu^-)$, dubbed ``true muonium''\cite{Hughes:1971}.  Simpler bound states like positronium $(e^+e^-)$, hydrogen, and muonium $(\mu^+e^-)$ have attracted significant attention as testing grounds for precision QED studies\cite{Karshenboim:2005iy}, but are limited in their BSM discovery potential by either the small electron mass or large uncertainties from unknown nuclear structure effects.  In contrast, true muonium has a much larger reduced mass, and its QCD corrections are limited to the better-understood hadronic vacuum polarization effects, due to its leptonic nature.  

Unfortunately, true muonium has yet to be directly observed.  The first reason is the technical difficulty of producing low-energy muon pairs, coupled with the bound state's short lifetime ($\tau\approx$ 1 ps), which presents an interesting challenge to experimenters.  Numerous proposed methods of production channels have been discussed over the years in the literature \cite{Nemenov:1972ph,Moffat:1975uw,Holvik:1986ty,Kozlov:1987ey,Ginzburg:1998df,ArteagaRomero:2000yh,Brodsky:2009gx,Chen:2012ci,Banburski:2012tk,Ellis:2015eea}, but until recently, none have been seriously developed.  A second, more prosaic, reason true muonium has been neglected is that, until the advent of the $(g-2)_\mu$ anomaly, there was no apparent reason to expect that true muonium would offer any novel physics compared to the simpler, more easily produced bound states.

Currently, the Heavy Photon Search (HPS)\cite{Celentano:2014wya} experiment has plans to search for true muonium \cite{Banburski:2012tk}, and DImeson Relativistic Atom Complex (DIRAC) \cite{Benelli:2012bw} has discussed the possibility of its observation in an upgraded run\cite{dirac}.  In both situations, the true muonium could be traveling at relativistic speeds, and it may be necessary to consider the effect of this boost on the wave functions \cite{Lamm:2013oga}.  Once observed, experiments focusing on precisely measuring the Lamb shift and hyperfine splitting (HFS) would put strong constraints on possible BSM solutions to the muon problem\cite{Heeck:2010pg,Jaeckel:2010xx,Batell:2011qq,Batell:2009jf,TuckerSmith:2010ra,Kopp:2014tsa,Gomes:2014kaa,Karshenboim:2010cm,Karshenboim:2010cg,Karshenboim:2010cj,Karshenboim:2010ck,Karshenboim:2011dx,Karshenboim:2014tka}.

Predictions for BSM models solving the muon problem generically lead to corrections to the spectrum of true muonium as large as $\mathcal{O}(100 $ MHz) (e.g, \cite{TuckerSmith:2010ra}).  This size corresponds to $\mathcal{O}(m\alpha^7)$ corrections to the true muonium spectrum; therefore, we must first have a standard model prediction of this level.  Currently, the theoretical predictions for the HFS in true muonium are known fully to $\mathcal{O}(m\alpha^5)$ \cite{Jentschura:1997tv,Jentschura:1997ma,Karshenboim:1998am}.  Additionally, all corrections of order $\mathcal{O}(m\alpha^6)$ (see \cite{Adkins:2014dva} and references within for a historical review) and partial results for order $\mathcal{O}(m\alpha^7)$ are known for positronium \cite{Melnikov:1999uf,Pachucki:1999zz,Kniehl:2000cx,Melnikov:2000zz,Hill:2000zy,Baker:2014sua,Adkins:2014dva,Eides:2014nga,Adkins:2014xoa}. With an exchange of the electron mass for the muon mass, these results can be included in the true muonium prediction, to yield the current theoretical value of
\begin{equation}
 \Delta E^{1s}_{\rm hfs}=42330685(800)\text{ MHz},
\end{equation}
where only the uncertainty arising from model-dependent hadronic effects\cite{Jentschura:1997tv} are included.

What remains of QED to be computed for true muonium at $\mathcal{O}(m\alpha^6)$ are corrections that do not occur in positronium involving virtual electrons and hadrons.  Without calculating all of these corrections, we would like some sense of the uncertainty, $\delta E^{6}_{\rm hfs}|_{\mu}$, in the theoretical value.  To do so we make the following, rather gross, approximation.  It is known that electron vacuum polarization corrections are the largest contribution from the work of \cite{Jentschura:1997tv,Jentschura:1997ma,Karshenboim:1998am} unique to true muonium at $\mathcal{O}(m\alpha^5)$.  We will use this fact to estimate the unknown $\mathcal{O}(m\alpha^6)$ corrections unique to true muonium to be the complete $\mathcal{O}(m\alpha^5)$ diagrams multiplied by the photon polarization function, $\Pi(q^2)$, that arises from the electron vacuum polarization at momentum $q^2=4m_\mu^2$.  While not perfect, the logic behind this estimate is that insertion of electron loops into photon propagators of $\mathcal{O}(m\alpha^5)$ diagrams constitute a large portion of the necessary corrections. This effect can estimated by the polarization function evaluated at the average expected momenta in the photon.  For single annihilation photons, this means $q^2\approx 4m_\mu^2$ photons, and $q^2\approx \alpha^2 m_\mu^2$ for exchange photons.  We will take the larger of these two.  A final point in our estimate is that, since the majority of $\mathcal{O}(m\alpha^5)$ contain two photon propagators, we should double the energy shift.
\begin{align}
 \delta E^{6}_{\rm hfs}|_{\mu}\approx& 2\Pi(4m_\mu^2)\Delta E^{5}_{\rm hfs}&\nonumber\\
 \approx&2\frac{\alpha}{\pi}\left[\frac{1}{3}\ln\left(\frac{4m_\mu^2}{m_e^2}\right)-\frac{5}{9}\right]\Delta E^{5}_{\rm hfs}\nonumber\\
 \approx&1200 \text{MHz},
\end{align}
where $\Delta E^{5}_{\rm hfs}$ was obtained in \cite{Jentschura:1997tv} and consists of the sum of the $\frac{m\alpha^5}{\pi}$ and $\frac{m\alpha^5}{\pi}|_{\mu}$ terms in Table~\ref{tab:1}. Regardless of the accuracy of this estimate, these terms must be computed to obtain $\mathcal{O}(100 $MHz$)$ precision.  

In this paper, we present a calculation of a novel contribution to the true muonium HFS that arises from the leading-order weak interactions, which scale as $\mathcal{O}(m\alpha^6)$, and a more general expression for the leading pseudovector contribution from a $Z'$ particle, which have previously been considered as a BSM source in other atomic systems \cite{Karshenboim:2010cm,Karshenboim:2010cg,Karshenboim:2010cj,Karshenboim:2010ck,Karshenboim:2011dx,Karshenboim:2014tka}.  

Previous work on weak interactions in electronic systems \cite{Beg:1974xv,Bodwin:1977ut,Bernreuther:1981ah,Kinoshita:1994ds,Eides:1995sq,Eides:2012ue} have consistently found the effects to be heavily suppressed.  This can be understood through the simple scaling associated with the weak interaction.  For electroweak couplings in atomic systems, the dimensionless coupling constant is given by $G_F \mu^2$ where $\mu$ is the reduced mass of the system.  In bound states, the $\mu^2$ term can be understood as arising partly from the wave function at the origin $|\psi(0)|^2=\mu^3\alpha^3/\pi n^3$ where $n$ is the energy level, and the remaining power of $\mu$ is needed for correct energy dimension.  Taking the known value of $G_F$, we see that the dimensionless constant can be expressed as $G_F \mu^2=0.097\mu^2/M_Z^2$.  From this result, an effective order of $\alpha$ can be computed for each atomic system via
\begin{equation}
 G_F \mu^2=0.097\frac{\mu^2}{M_Z^2}=\alpha^n.
\end{equation}
The leading-order electroweak correction to electronic systems should then be proportional to $G_F |\psi(0)|^2\simeq G_F m_e^3\alpha^3\simeq m_e\alpha^{8.4}$.  The theoretical uncertainties for the positronium, muonium, and hydrogen states are all far larger than this.

This situation should be compared with that in true muonium and muonic hydrogen, where the coupling instead is proportional to $G_F m_\mu^2\simeq \alpha^{3.2}$, leading to a naive prediction of its effect on the HFS of $\simeq$$m_\mu\alpha^{6.2}$.  In muonic hydrogen, this effect will be difficult to discern because of the discrepancy in the proton radius and the unknown nuclear effects that plague hydrogenic atoms.  In contrast, the almost purely QED atom of true muonium should offer a cleaner opportunity to measure these electroweak effects.  As stated above, to see these effects, a number of QED contributions of $\mathcal{O}(m\alpha^6)$ and $\mathcal{O}(m\alpha^7)$ need to be computed. Putting these contributions together, a prediction for the HFS of true muonium to 1 MHz precision should be possible.

In this work, we extend the previous work in \cite{Eides:1995sq} by computing the leading-order energy shifts to HFS from the exchange and annihilation of a pseudovector in QED particle-antiparticle bound states, with special emphasis on true muonium, the most viable candidate in which to measure these effects in the near future.  Numerical values using the $Z$ boson are reported.  In Sec.~\ref{sec:2} we review the previous calculations for the exchange process, and in Sec.~\ref{sec:3} we compute the results for the annihilation channel. We devote Sec.~\ref{sec:4} to discussing the total contribution, and conclude in Sec.~\ref{sec:5}.

\section{Exchange of $Z$ boson}
\label{sec:2}
Calculations have existed for many years for the absolute value of the contribution from the $t$-channel exchange of a single $Z$ boson to atomic systems \cite{Beg:1974xv,Bodwin:1977ut,Kinoshita:1994ds}, but the correct sign for this effect was unclear.  In \cite{Eides:1995sq}, the author resolved the issue, with the conclusion that the effect in hydrogen is positive, while in muonium it is negative due to the antiparticle nature of $\mu^+$.  In \cite{Eides:2012ue} the HFS for other light muonic atoms were calculated, but the case of true muonium was conspicuously left untouched.  Following \cite{Eides:1995sq}, we write the leading-order Lagrangian for the neutral weak interaction as
\begin{equation}
 L_{int}=\bar{g}\bar{L}\gamma_\mu T_3 L Z^\mu,
\end{equation}
where $\bar{g}=g/\cos\theta_W$, and the weak left-handed spinor isodoublet $L$ is defined using the convention:
\begin{equation}
 L=\frac{1+\gamma_5}{2}\psi.
\end{equation}
While the $Z$ boson has a vector component, this is further suppresed by a factor $1-4\sin^2\theta_W$. Therefore we will neglect it's contribution in this work.  Generalizing the exchange correction result of \cite{Eides:1995sq} for any particle-antiparticle bound state to the HFS, we obtain
\begin{equation}
 \Delta E^x_{\rm hfs}(n)=-\frac{2G_F}{\sqrt{2}}\frac{(Z\alpha)^3}{\pi n^3}\mu^3,
\end{equation}
where $Z$ is the charge of the particle.  Replacing the reduced mass by $m_l/2$ and taking $Z=1$, we have that the correction to lepton-antilepton bound states is
\begin{equation}
\label{eq:exch}
\Delta E^x_{l,\rm hfs}(n)=-\frac{G_F}{4\sqrt{2}}\frac{m_l^3\alpha^3}{\pi n^3}.
\end{equation}
In a similar way, we may write the energy shift for a new $Z'$ boson in the exchange channel as
\begin{equation}
\label{eq:exch1}
 \Delta E^{x,Z'}_{l,\rm hfs}(n)=-\frac{g^2}{32M_{Z'}^2}\frac{m_l^3\alpha^3}{\pi n^3},
\end{equation}
where $g$ is the new axial vector coupling between $Z'$ and the leptons.

\begin{table*}[t]
\begin{center}
 \begin{tabular}{c c c c}
 \hline\hline
  $\mathcal{O}(m\alpha^n)$ & $C_n$ & $\delta E^{1s}_{\rm hfs}  \rm [MHz]$&Ref.\\
  \hline
   $m\alpha^4$ & $\frac{7}{12}$	&42260692	&  \cite{Pirenne:1947,Berestetskii:1949,Ferrell:1951zz}\\
   $\frac{m\alpha^5}{\pi}$ & $-\frac{1}{2}\ln(2)-\frac{8}{9}$	&-207904	& \cite{Karplus:1952wp} \\
   $\frac{m\alpha^5}{\pi}|_{\mu}$ &1.638(5) &275572(800)	& \cite{Jentschura:1997tv} \\
   $\frac{m\alpha^6}{\pi^2}$ &$-\frac{52}{32}\zeta(3)+\left(\frac{221}{24}\ln(2)-\frac{5197}{576}\right)\zeta(2)+\frac{1}{2}\ln(2)+\frac{1367}{648}$	&-1515	& \cite{Czarnecki:1998zv} \\
   $\frac{m\alpha^6}{\pi^2}|_{\mu}$ &--- &(1200)	&Estimated  \\
   $m\alpha^6\ln\left(\frac{1}{\alpha}\right)$ &$\frac{5}{24}$	&3954	&  \cite{Lepage:1977gd}\\
   $\frac{m\alpha^7}{\pi^3}$ &	158*&144*	& \cite{marcu2011ultrasoft,Baker:2014sua,Adkins:2014dva,Eides:2014nga,Adkins:2014xoa} \\
   $\frac{m\alpha^7}{\pi}\ln\left(\frac{1}{\alpha}\right)$ &$-\frac{17}{3}\ln(2)+\frac{217}{90}$	&-67	& \cite{Kniehl:2000cx,Melnikov:2000zz,Hill:2000zy} \\
   $\frac{m\alpha^7}{\pi}\ln^2\left(\frac{1}{\alpha}\right)$ &$-\frac{7}{8}$	&-190	& \cite{Karshenboim:1993} \\
   $\frac{m^3G_F\alpha^3}{\sqrt{2}\pi}$&$-\frac{3}{8}$	&-109	&This work\\
   $\frac{m\alpha^7}{\pi^2}|_{\mu}$ &--- &---	&---  \\
   \hline
   Total &	&42330577(800)(1200)	&\\
  \hline\hline
 \end{tabular}
\end{center}
 \caption{\label{tab:1}Predictions for the $1s$ HFS in true muonium.  The designation of $|_\mu$ indicates corrections for true muonium missing from positronium, which depend upon $m_\mu/m_e$.  The error in the table consists of two estimated parts, the former from the model dependence of the hadronic loops, and the latter from electronic loop effects at $\mathcal{O}(m\alpha^6)$, as described in text. $*$ indicate terms that are only partially computed.}
\end{table*}
\section{Annihilation via virtual $Z$ boson}
\label{sec:3}
For the true muonium and positronium systems, there is an additional diagram coming from the $Z$ boson annihilation channel.  Following \cite{Eides:1995sq}, we find that the scattering amplitude for for this process is given by
\begin{equation}
 \frac{\bar{g}^2T_3(2)T_3(1)}{4}\frac{\bar{v}(2)\gamma^5\gamma^\mu u(1)\bar{u}(1)\gamma_{\mu}\gamma^5 v(2)}{(2m)^2-M_z^2},
\end{equation}
where $T_3(i)$ is the weak isospin of particle or antiparticle $i$.  An additional minus sign occurs due to the relative sign in the weak neutral current of the antiparticle compared to the particle.  This expression can be Fierz reordered to yield
 \begin{align}
 \frac{\bar{g}^2T_3(2)T_3(1)}{16[(2m)^2-M_z^2]}&\times\nonumber\\\times \big[4\bar{v}(2)v(2)\bar{u}(1)&u(1)+2\bar{v}(2)\gamma^{\mu}v(2)\bar{u}(1)\gamma_{\mu}u(1)\nonumber\\-2\bar{v}(2)\gamma^{5}\gamma^{\mu}v&(2)\bar{u}(1)\gamma_{\mu}\gamma^{5}u(1)-4\bar{v}(2)\gamma^{5}v(2)\bar{u}(1)\gamma^{5}u(1)\big].
 \end{align}
To simplify this expression, we take the non-relativistic limit ($\bm{p}_i\rightarrow 0$), divide by the relativistic normalization $\sqrt{2m}$ for each spinor, and integrate over an $s$-state wave function to get the wave function squared at origin, multiplied by constants.  It is important to note that for pseudovectors with $M_{Z'}\leq m_l$, there are corrections that arise from the Yukawa-like interaction that depend on $M_{Z'}r$ where $r$ is the radius of the atom.  We will take the assignments of weak isospin for the antiparticle to be $T_3(2)=-1/2$ and for the particle to be $T_3(1)=1/2$.  From this, we have the expression:
\begin{equation}
 \langle \psi|\Delta H_{Z}|\psi \rangle =\frac{\bar{g}^2}{16[(2m)^2-M^2_Z]}|\psi(0)|^2\bigg<\frac{1-\bm{\sigma}_2\cdot\bm{\sigma}_1}{2}\bigg>.
\end{equation}
Noting that $\langle S^2\rangle =\langle\frac{3+\bm{\sigma}_2\cdot\bm{\sigma}_1}{2}\rangle$, we can rewrite the expectation value of the spins.  Using the non-relativistic value for $|\psi(0)|^2$ given in Sec.~\ref{sec:1}, and evaluating this Hamiltonian for the difference between the singlet and triplet state, $\langle 2-\bm{S}^2\rangle=2$, we have 
\begin{equation}
\label{eq:anni1}
  \Delta E^a_{\rm hfs}=\frac{\bar{g}^2}{8[(2m)^2-M^2_Z]}\frac{\mu^3\alpha^3}{\pi n^3}.
\end{equation}
For the case of a general $Z'$ boson, this is the final expression to consider after appropriate variable exchanges.  It is interesting to note that for $M_{Z'}\approx(2m)^2$, this term will dominate over the exchange term.  This may potentially lead to stronger constraints on BSM pseudovectors for this mass range than those found for positronium in \cite{Karshenboim:2010cg}.  In contrast, for the real electroweak $Z$ boson, we can safely approximate $(2m)^2-M^2_Z\approx -M^2_Z$.  Using the definition that $\bar{g}$ from Sec.~\ref{sec:2} and the relations $g^2=4\pi\alpha/\sin^2\theta_W$, we have
\begin{equation}
 \Delta E^a_{\rm hfs} =-\frac{1}{8}\frac{4\pi\alpha}{\sin^2\theta_W\cos^2\theta_WM^2_Z}\frac{\mu^3\alpha^3}{\pi n^3}.	
\end{equation}
Using the relation between the massive boson masses $M_W=M_Z\cos(\theta_W)$ gives
\begin{equation}
 \Delta E^a_{\rm hfs} =-\frac{1}{8}\frac{4\pi\alpha}{\sin^2\theta_WM^2_W}\frac{\mu^3\alpha^3}{\pi n^3}.	
\end{equation}
Finally, using the relation $G_F/\sqrt{2}=4\pi\alpha/8M^2_{W}\sin^2(\theta_w)$, and that the lepton mass is related to the reduced mass by $\mu=m_l/2$ we reach 
\begin{equation}
\label{eq:anni}
 \Delta E^a_{\rm hfs} =-\frac{G_F}{8\sqrt{2}}\frac{m_l^3\alpha^3}{\pi n^3}.	
\end{equation}
Comparing this to Eq.~(\ref{eq:exch}), we see that the annihilation channel increases the electroweak correction by 50\%. 

\section{Total Contribution}
\label{sec:4}
Summing Eq.~(\ref{eq:exch1}) and Eq.~(\ref{eq:anni1}), we find that the total pseudovector contribution to the hyperfine splitting of true muonium or positronium is 
\begin{equation}
 \Delta E^{Z'}_{l,\rm hfs}(n)=\left[-\frac{g^2}{4M_{Z'}^2}+\frac{g^2}{8[(2m)^2-M^2_{Z'}]}\right]\frac{m_l^3\alpha^3}{8\pi n^3}.
\end{equation}
Taking instead Eq.~(\ref{eq:exch}) and Eq.~(\ref{eq:anni}) for the specific case of the electroweak $Z$ boson:
\begin{equation}
 \Delta E^Z_{l,\rm hfs} =-\frac{3G_F}{8\sqrt{2}}\frac{m_l^3\alpha^3}{\pi n^3}.
\end{equation}
Using our approximate value of $G_Fm_\mu^2\simeq \alpha^{3.2}$, we find that as predicted, this correction does naively scale as $\mathcal{O}(m_\mu\alpha^6)$.  Closer inspection of the actual numerical coefficient reveals that this contribution is comparable in size to the $\mathcal{O}(m_\mu\alpha^7)$.  For these particle-antiparticle bound states, the leading-order HFS is given by $\Delta E^{(0)}_{\rm hfs}=\frac{7}{12}\alpha^4m_l$.  As a fraction of the leading order, we have corrections to true muonium of the size
\begin{equation}
 \Delta E^Z_{\mu,\rm hfs}=-2.58\times10^{-6}E^{(0)}_{\mu,\rm hfs},
\end{equation}
and that the the total shift is $-109$ MHz.  This correction has been included in Table~\ref{tab:1}, which gives the current state of the theoretical predictions.

\section{Conclusion}
\label{sec:5}
In conclusion, upcoming experiments present for the first time the strong possibility to detect and measure the properties of true muonium.  These results will allow for strong discrimination of BSM explanations for the muon problem.  To theoretically differentiate these models, one will need a precision of $\mathcal{O}(100$ MHz) for the hyperfine splitting of true muonium.  

In this paper, we have presented a subleading, but perhaps surprisingly large, correction from the electroweak interaction that needs to be accounted for in order to achieve the required precision.  We have additionally computed the leading-order correction due to a general $Z'$ boson.  Further work remains to compute the more standard QED effects to higher order and to dramatically reduce the model-dependent uncertainties in the hadronic loops.
\begin{acknowledgments}
The author would like to thank Richard Lebed and Jayden Newstead for enlightening comments while developing this work.  This work was supported by the National Science Foundation under Grant Nos. PHY-1068286 and PHY-1403891.
\end{acknowledgments}
\bibliographystyle{apsrev4-1}
\bibliography{wise}
\end{document}